\documentclass[twocolumn,superscriptaddress,amsmath,amssymb,longbibliography]{revtex4-1}
\usepackage[pdftex]{graphicx}
\usepackage{dcolumn}
\usepackage{here}

\begin{document}

\title{Emergence of low-energy electronic states in oxygen-controlled Mott insulator Ca$_{2}$RuO$_{4+\delta}$}

\author{Takeo Miyashita}
\affiliation{%
Graduate School of Science, Hiroshima University, 1-3-1 Kagamiyama, Higashi-Hiroshima, Hiroshima 739-8526, Japan
}%

\author{Hideaki Iwasawa}
\affiliation{%
Quantum Beam Science Research Directorate, National Institutes for Quantum and Radiological Science and Technology, 1-1-1 Koto, Sayo, Hyogo 679-5148, Japan;
}%

\author{Tomoki Yoshikawa}
\affiliation{%
Graduate School of Science, Hiroshima University, 1-3-1 Kagamiyama, Higashi-Hiroshima, Hiroshima 739-8526, Japan
}%

\author{Shusuke Ozawa}
\affiliation{%
Graduate School of Science, Hiroshima University, 1-3-1 Kagamiyama, Higashi-Hiroshima, Hiroshima 739-8526, Japan
}%

\author{Hironoshin Oda}
\affiliation{%
Graduate School of Science, Hiroshima University, 1-3-1 Kagamiyama, Higashi-Hiroshima, Hiroshima 739-8526, Japan
}%

\author{Takayuki Muro}
\affiliation{%
Japan Synchrotron Radiation Research Institute, 1-1-1 Kouto, Sayo, Hyogo 679-5198, Japan
}%

\author{Hiroki Ogura}
\affiliation{%
Kurume Institute of Technology, 2228-66 Kamitsu, Kurume, Fukuoka 830-0052, Japan
}%

\author{Tatsuhiro Sakami}
\affiliation{%
Kurume Institute of Technology, 2228-66 Kamitsu, Kurume, Fukuoka 830-0052, Japan
}%

\author{Fumihiko Nakamura}
\affiliation{%
Kurume Institute of Technology, 2228-66 Kamitsu, Kurume, Fukuoka 830-0052, Japan
}%

\author{Akihiro Ino}
\email{ino@kurume-it.ac.jp}
\affiliation{%
Kurume Institute of Technology, 2228-66 Kamitsu, Kurume, Fukuoka 830-0052, Japan
}%
\affiliation{%
Hiroshima Synchrotron Radiation Center, Hiroshima University, 2-313 Kagamiyama, Higashi-Hiroshima, Hiroshima 739-0046, Japan
}%

\date{\today}

\begin{abstract}
Insulator-to-metal transition in Ca$_{2}$RuO$_{4}$ has drawn keen attention because of its sensitivity to various stimulation and its potential controllability.
Here, we report a direct observation of Fermi surface, which emerges upon introducing excess oxygen into an insulating Ca$_{2}$RuO$_{4}$, by using angle-resolved photoemission spectroscopy.
Comparison between energy distribution curves shows that the Mott insulating gap is closed by eV-scale spectral-weight transfer with excess oxygen.
Momentum-space mapping exhibits two square-shaped sheets of the Fermi surface.
One is a hole-like $\alpha$ sheet around the corner of a tetragonal Brillouin zone, and the other is an electron-like $\beta$ sheet around the $\Gamma$ point.
The electron occupancies of the $\alpha$ and $\beta$ bands are determined to be $n_{\alpha}=1.6$ and $n_{\beta}=0.6$, respectively. 
Our result indicates that the insulator-to-metal transition occurs selectively in $d_{xz}$ and $d_{yz}$ bands and not yet in $d_{xy}$ band.
This orbital selectivity is most likely explained in terms of the energy level of $d_{xy}$, which is deeper for Ca$_{2}$RuO$_{4+\delta}$ than for Ca$_{1.8}$Sr$_{0.2}$RuO$_{4}$.
Consequently, we found substantial differences from the Fermi surface of other ruthenates, shedding light on a unique role of excess oxygen among the metallization methods of Ca$_{2}$RuO$_{4}$.
\end{abstract}

\maketitle

\section{Introduction}

Needs for controlling phase transitions and physical properties have been a driving force toward the development of material science.
This has driven researchers' attention to Mott insulators, where strong Coulomb repulsion forces the electrons to be localized despite a sufficient number of electrons~\cite{Mott1968,Imada1998}, and thus the metallic conductivity can be triggered by small perturbation to the insulating state.
Furthermore, in the vicinity of the Mott transition, striking phenomena such as colossal magnetoresistance in manganites~\cite{Tokura1994} and high-critical-temperature superconductivity in cuprates~\cite{Bednorz1986} have been discovered.\\
\hspace*{2ex}One of the most dramatic Mott transitions is known to occur in a layered ruthenate, Ca$_{2}$RuO$_{4}$.
The ground state of Ca$_{2}$RuO$_{4}$ is an insulating phase.
With increasing temperature, the transition to a metallic phase occurs at $T_\mathrm{MI} = 364$~K along with a structural transition~\cite{Alexander1999}.
The colossal negative thermal expansion observed below $T_\mathrm{MI}$~\cite{Takenaka2017} also implies the role of structural distortion in this transition.
In addition to the temperature, the insulator-to-metal transition is caused by applying pressure~\cite{Nakamura2002,Nakamura2007}, elemental substitution for Ca~\cite{Nakatsuji2000_1,Nakatsuji2000_2}, and introducing excess oxygen~\cite{Braden1998} for Ca$_{2}$RuO$_{4}$.
Remarkably, it has been found that the transition to the metallic phase is realized by applying an electric field as weak as 40 V/cm for Ca$_{2}$RuO$_{4}$~\cite{Nakamura2013}, suggesting that this transition is controllable on the device application.
Beyond the metallic phase, superconducting phases are also recognized for Ca$_{2}$RuO$_{4}$ under high pressure~\cite{Alireza2010}, and for Sr$_{2}$RuO$_{4}$ with full substitution of Sr for Ca~\cite{Maeno2011,Mackenzie2017}.
In order to manipulate the phase transitions in Ca$_{2}$RuO$_{4}$, the concrete shape of Fermi surface in the metallic phase would provide fundamental basis for a microscopic picture.\\
\hspace*{2ex}However, it is increasingly controversial which Fermi surface is directly involved in the insulator-to-metal transition of Ca$_{2}$RuO$_{4}$.
Since the electron configuration of Ru $4d$ orbitals is $d^4$, four $d$ electrons per Ru enter in three $t_\mathrm{2g}$ bands~\cite{Sutter2017}.
According to the standard notation for Sr$_{2}$RuO$_{4}$, we refer to the lower and upper branches of the hybridization of the $d_{xz}$ and $d_{yz}$ bands as $\alpha$ and $\beta$, respectively, and the $d_{xy}$ band as $\gamma$~\cite{Mackenzie1996}.
Early calculation using dynamical mean-field theory proposed that the insulator-to-metal transition occurs selectively in the $d_{xy}$ band~\cite{Anisimov2002}.
After that, the Fermi surface induced by 10\% substitution of Sr for Ca and by uniaxial strain has been reported by angle-resolved photoemission spectroscopy (ARPES), and some critical discrepancies have been raised~\cite{Shimoyamada2009,Neupane2009,Kim2020,Ricco2018}.
Specifically, while the $\alpha$ sheet of Fermi surface is commonly observed by ARPES, it is under debate whether or not the $\beta$ and/or $\gamma$ sheets of Fermi surface emerge upon the transition.
In addition, the scenarios for the evolution of the electronic structure across the insulator-to-metal transition are complicated by the band folding due to orthorhombic structural distortion, which is weakened but usually persists in the metallic phase of Ca$_{2}$RuO$_{4}$.
Therefore, it is interesting to explore the Fermi surface induced by an alternative method of metallization, so that we may find out the features inherent to the Mott transition in Ca$_{2}$RuO$_{4}$.

It has been recognized that the Mott insulating phase of Ca$_{2}$RuO$_{4}$ can be suppressed by the presence of non-stoichiometric excess oxygen~\cite{Braden1998,Takenaka2017}.
Indeed, vacancy and excess of oxygen play a key role in a resistive switching phenomenon in the devices based on transition-metal oxides~\cite{Sawa2008} and in the electric-field control of the Mott transition in VO$_{2}$~\cite{Jeong2013}.
As is usual with transition-metal oxides, the dependence on the actual amount of oxygen has been a potential source of confusions and discrepancies between experiments of ruthenates.
For these reasons, direct observation of the electronic structure of oxygen-controlled Ca$_{2}$RuO$_{4+\delta}$ is expected to make new insights into the mechanism and control of the Mott transition. 

\begin{figure}[!t]
\begin{center}\includegraphics[width=\columnwidth]{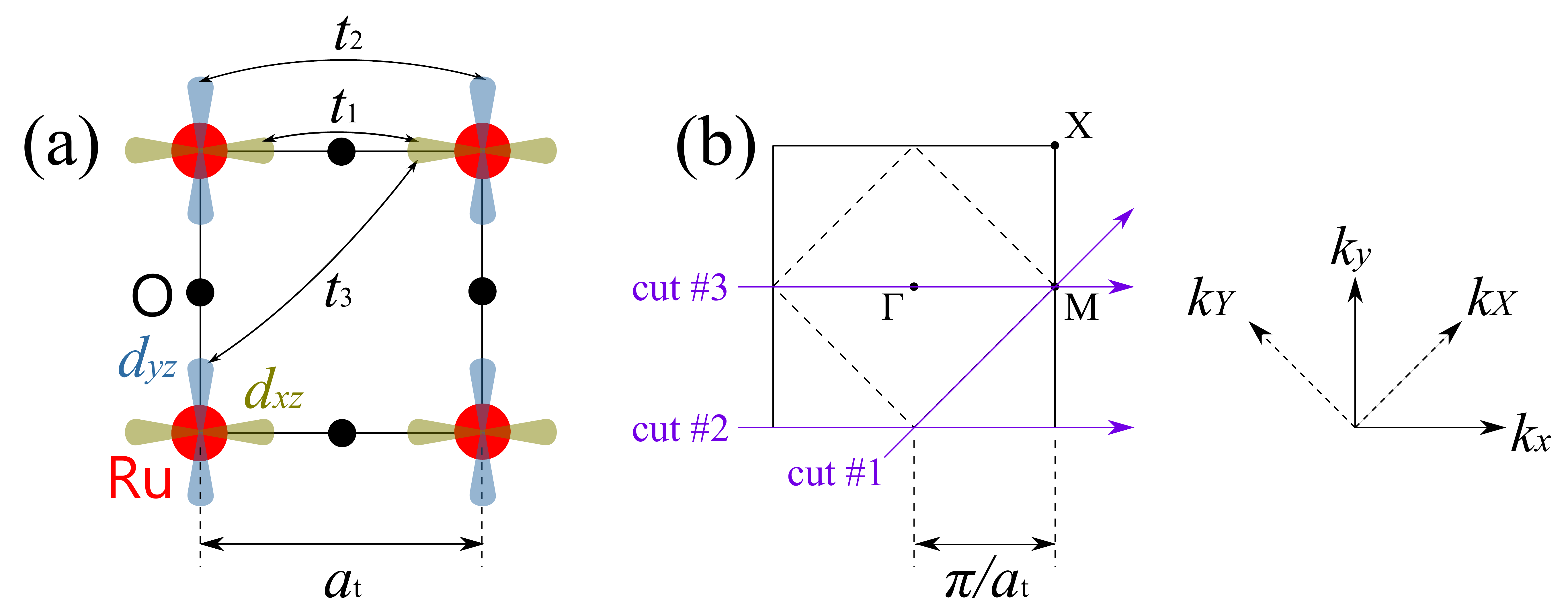}
\caption{
(a) Tetragonal unit of RuO$_{2}$ plane in real space.
Relevant $d_{xz}$ and $d_{yz}$ orbitals are superimposed on Ru atoms.
(b) Two-dimensional view of momentum space.
Solid and dashed lines represent the boundaries of tetragonal and orthorhombic Brillouin zones (BZs), respectively.
Momentum points, M and X, are defined so that $\Gamma$-M direction is along tetragonal axes, $k_{x}$ and $k_{y}$ (solid arrows), and that $\Gamma$-X direction is along orthorhombic axes, $k_{X}$ and $k_{Y}$ (dashed arrows).
}
\label{fig1}
\end{center}
\end{figure}

  %

Here, we report the first ARPES study of the metallic electronic states induced by excess oxygen for Ca$_{2}$RuO$_{4+\delta}$, taking advantage of high-quality single-crystalline samples and a micro-spot soft X-ray beam, which enables us to select a broadly uniform region of the sample surface.
We present a stark contrast in low-energy electronic states between stoichiometric and excess-oxygen samples.
On the basis of experimental image of the Fermi surface, we address the orbital character of the electrons directly responsible for the insulator-to-metal transition in Ca$_{2}$RuO$_{4+\delta}$, and then consider possible differences among the multiple metallization methods.

\section{Experiment}

High-quality single crystals of Ca$_{2}$RuO$_{4+\delta}$ were grown by a traveling-solvent floating-zone method with RuO$_{2}$ self-flux as described in Ref.~\citenum{Nakamura2013}.
The amount of excess oxygen, $\delta$, was controlled by the mixing ratio of argon-oxygen gas filled in a quartz tube during the crystal growth. 
While the total pressure of mixture gas was 10 atm, the partial pressure of oxygen gas was 1 and 3 atm for stoichiometric and excess-oxygen samples, respectively.
The insulator-to-metal transition has been observed at $T_\mathrm{MI} = 364$~K and 358~K for stoichiometric and excess-oxygen samples, respectively, by heat-capacity measurement.
Soft X-ray ARPES measurements were performed at BL25SU of SPring-8 using circularly-polarized synchrotron radiation. 
The spot size of the incident light is about 10 $\mu$m $\times$ 13 $\mu$m on the sample surface.
All the ARPES spectra were recorded with a DA30 electron analyzer (Scienta-Omicron) equipped with deflector lens.
The deflector function enables to measure the two-dimensional angular distribution of photoelectrons without mechanical sample/analyzer rotation while keeping the experimental geometry.
The total energy resolution was set at 60 meV for a photon energy of 400 eV.
Clean surfaces were obtained by \textit{in situ} cleaving at a pressure lower than $1\times 10^{-7}$ Pa, and the samples were kept at room temperature of 302 K, which is lower than $T_\mathrm{MI}$, during the ARPES experiment.
The energy was calibrated with respect to the Fermi edge of a polycrystalline gold, and expressed as relative to the Fermi level, $E_\mathrm{F}$, in this paper.

Figure~\ref{fig1} illustrates the real and momentum spaces of two-dimensional RuO$_2$ plane.
If the tilting and rotation of RuO$_6$ octahedron are ignored, the crystal system would be tetragonal as in Sr$_2$RuO$_4$.
The tetragonal lattice constant, $a_\mathrm{t} \simeq 3.85$ \AA, is given by the nearest Ru-Ru distance, as shown in Fig.~\ref{fig1}(a).
In reality, due to the deformation of the RuO$_2$ plane, the orthorhombic unit cell is doubled in the real space.
Correspondingly, the Brillouin zone (BZ) is folded to half in the momentum space.
The boundaries of the tetragonal and orthorhombic BZs are denoted by solid and dashed lines, respectively, in Fig.~\ref{fig1}(b).
Here, we define $k_{x}$-$k_{y}$ and $k_{X}$-$k_{Y}$ coordinate axes along the tetragonal and orthorhombic reciprocal primitive vectors, respectively.
In this paper, we adopt the tetragonal notation for the labels of high-symmetry points, X and M, as shown in Fig.~\ref{fig1}(b), because the unfolded BZ is convenient to describe our experimental results.
We also use momentum units of $\pi/a_\mathrm{t} = 0.816$ \AA$^{-1}$ and $\pi/(\sqrt{2}a_\mathrm{t}) = 0.577$ \AA$^{-1}$ to display ARPES images.

\begin{figure}[t]
\begin{center}
\includegraphics[width=\columnwidth]{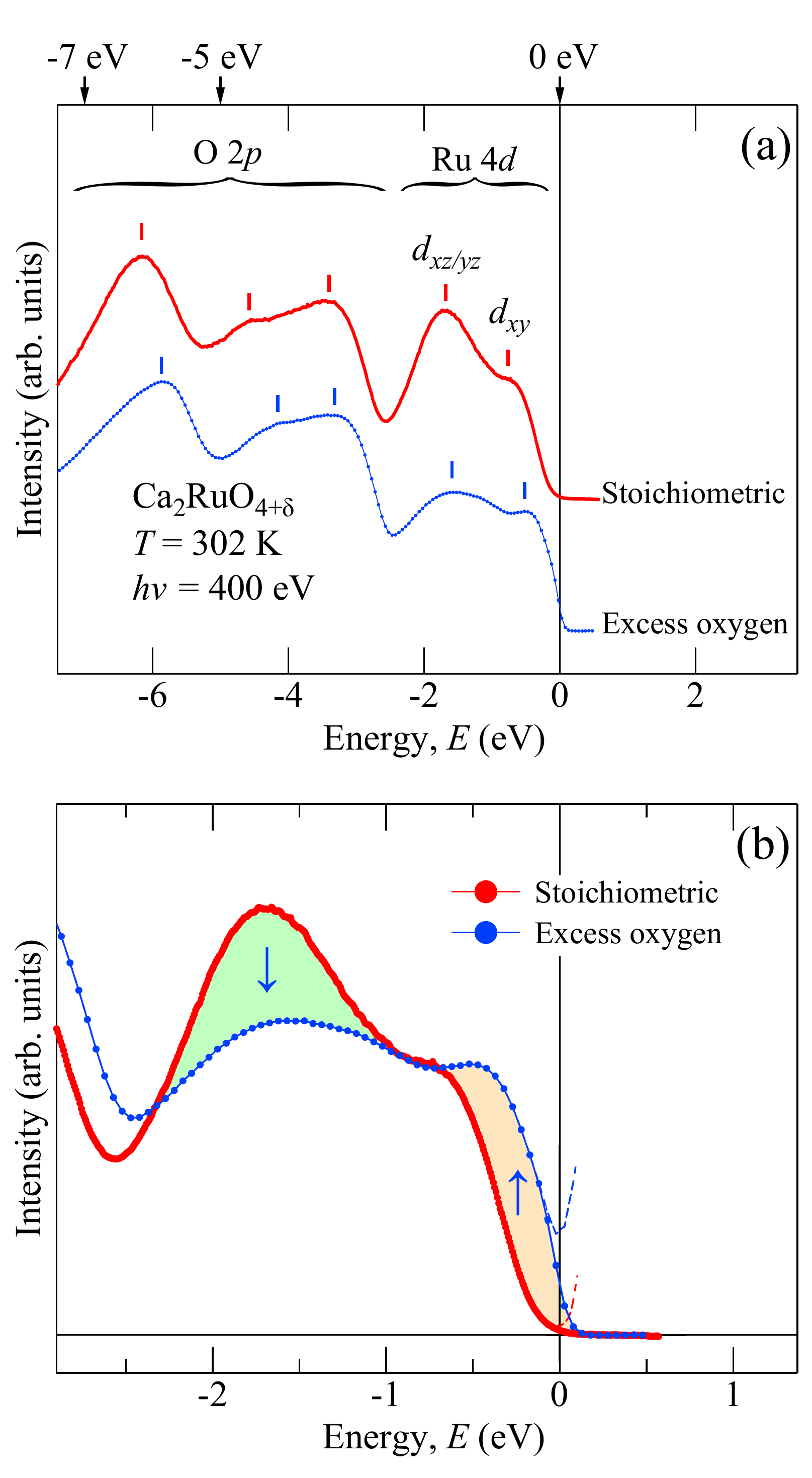}
\caption{Photoemission energy distribution obtained by integrating ARPES spectra over angles for Ca$_{2}$RuO$_{4+\delta}$.
Red and blue denote stoichiometric and excess-oxygen samples, respectively.
The intensity was normalized to the spectral area for $E > -2.6$~eV.
(a) Wide-range view of valence band.
(b) Enlarged view around the Fermi level.
The spectral difference is highlighted by green and orange.
The spectra divided by Fermi-Dirac distribution function are shown as dashed curves.
}
\label{fig2}
\end{center}
\end{figure}

\begin{figure}[t]
\begin{center}
\includegraphics[width=\columnwidth]{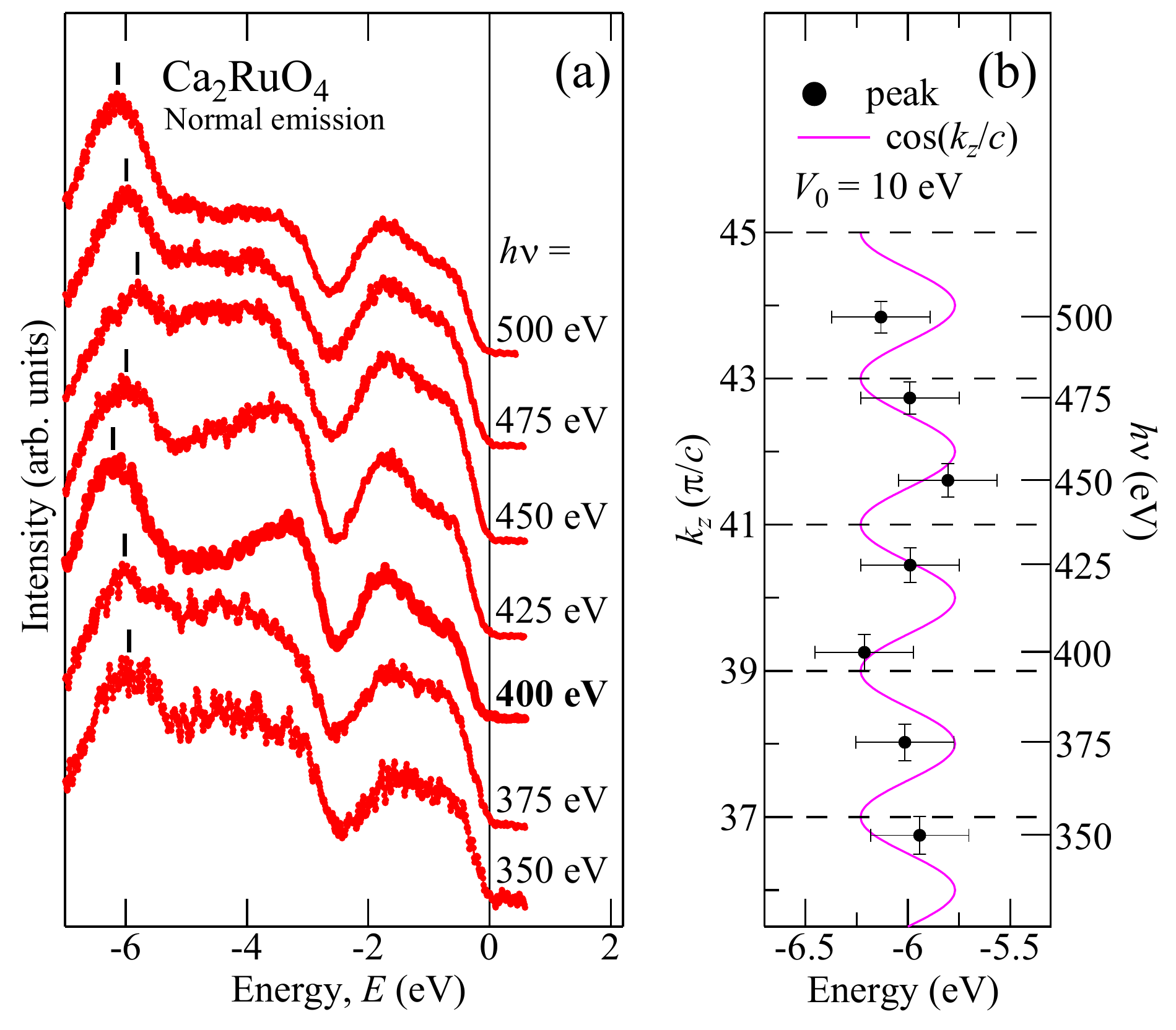}
\caption{
(a) Photon-energy dependence of normal-emission spectrum of stoichiometric Ca$_{2}$RuO$_{4}$.
(b) Energy of spectral peak at $\sim -6$~eV (filled circles), compared with a periodic function, $\cos(k_{z}/c)$ (magenta curve).
Photon energy, $h\nu$ (right axis), is related to perpendicular momentum, $k_{z}$ (left axis), on the assumption of an inner potential of $V_{0} = 10$~eV. 
}
\label{fig3}
\end{center}
\end{figure}

\begin{figure*}[!t]
\begin{center}
\includegraphics[width=15 cm]{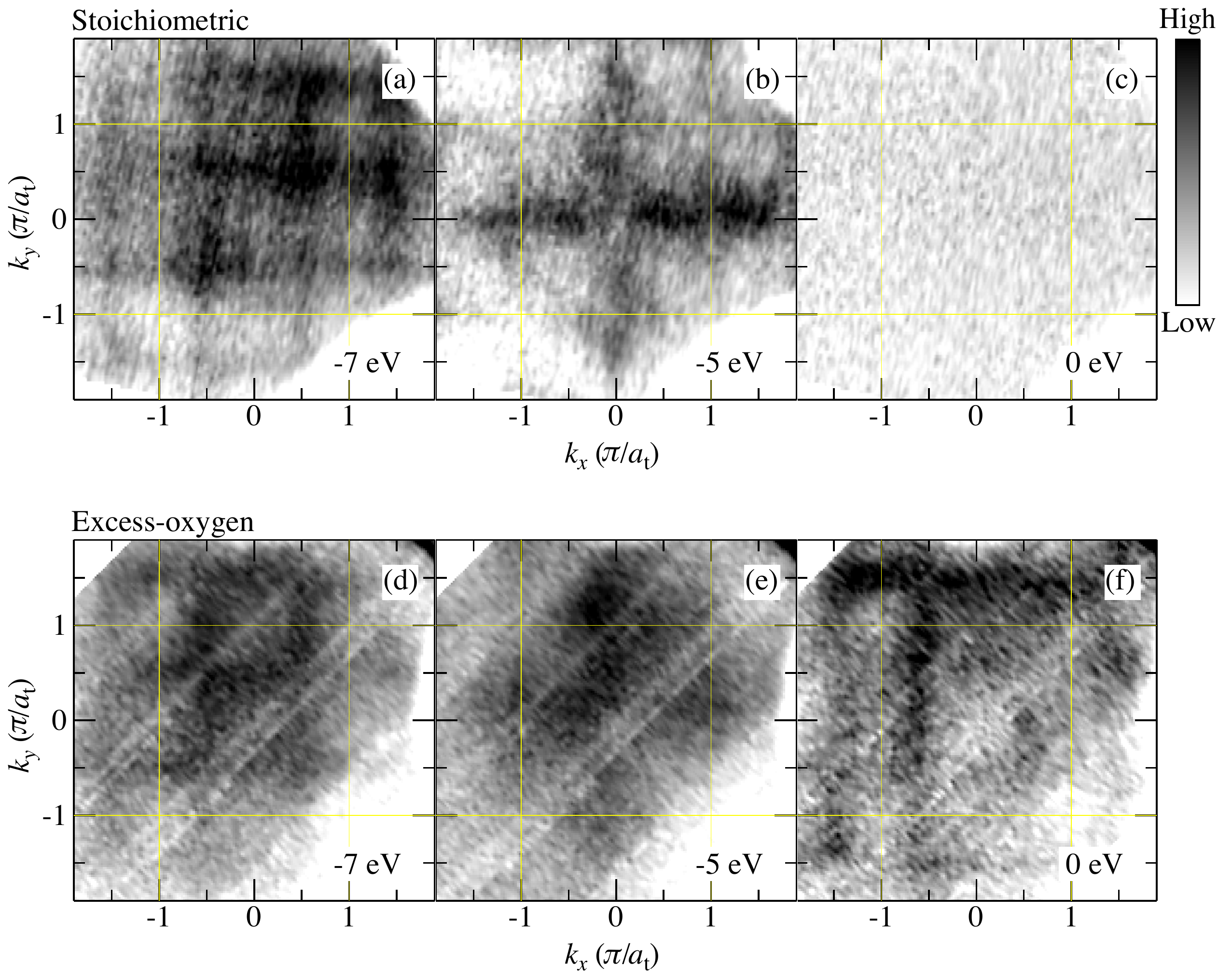}
\caption{
Momentum-space mappings at representative energies.
The values of momenta are given in unit of $\pi/a_\mathrm{t}$, and the tetragonal zone boundaries are marked by thin yellow lines.
The gray-scale varies from white to black following the low-to-high intensity.
Upper panels: data taken for a stoichiometric sample at energies of (a) $-7$~eV, (b) $-5$~eV, and (c) 0 eV.
Lower panels: data taken for an excess-oxygen sample at energies of (d) $-7$~eV, (e) $-5$~eV, and (f) 0 eV.
}
\label{fig4}
\end{center}
\end{figure*}

\section{Result}

A clear-cut overview of electronic density of states (DOS) of Ca$_{2}$RuO$_{4}$ has been obtained as shown in Fig.~\ref{fig2}(a). 
The energy distribution curves were obtained by integrating ARPES spectra over emission angles, and plotted with red and blue markers for stoichiometric and excess-oxygen samples, respectively. 
Valence-band peaks are observed at $E = -6.1$, $-4.6$, $-3.4$, $-1.7$, and $-0.8$~eV for stoichiometric Ca$_{2}$RuO$_{4}$.
They are divided by the orbital character.
The electronic states below and above an energy of $E = -2.6$~eV mainly have characters of O $2p$ and Ru $4d$ orbitals, respectively.
The latter is of our interest, as enlarged in Fig.~\ref{fig2}(b), because the lower Hubbard band near $E_\mathrm{F}$ is responsible for the insulator-to-metal transition.
We assigned the peaks at $-1.7$ and $-0.8$~eV to $d_{xz/yz}$ and $d_{xy}$ bands, respectively, according to the mean-field calculation compared with a previous ARPES experiment~\cite{Sutter2017}.
It can clearly be seen from Figs.~\ref{fig2}(a) and (b) that no spectral intensity is observed at $E_\mathrm{F}$ for stoichiometric sample, indicating that the Mott insulating gap is open.
The gap opening is consistent with the transport studies showing that the stoichiometric Ca$_{2}$RuO$_{4}$ is in Mott insulating state at room temperature~\cite{Nakatsuji1997, Braden1998, Nakatsuji2000_2}.

Impact of introducing excess oxygen into Ca$_{2}$RuO$_{4}$ is substantial.
Figure~\ref{fig2}(a) shows that the spectral features of O $2p$ band shift by about 0.2 eV towards $E_\mathrm{F}$.
The direction of the energy shift is consistent with the hole doping that is expected as the standard effect of excess oxygen.
Figure~\ref{fig2}(b), however, shows that a change in Ru $4d$ band is not rigid-band-like.
As excess oxygen is introduced, the peak intensity of $d_{xz/yz}$ band considerably decreases (green), and finite spectral weight appears at $E_\mathrm{F}$ (orange), raising questions about the orbital character and momentum distribution of the electronic states at $E_\mathrm{F}$.
The spectrum divided by a Fermi-Dirac distribution function indicates that the spectral weight at $E_\mathrm{F}$ is as much as 40\% of the height of leading edge for the excess-oxygen samples.
Even though the amount of excess oxygen is small, the spectral-weight transfer occurs in eV scale, and results in the closing of the Mott insulating gap.
This suggests that the transition to the metallic state has been caused by the excess oxygen within the photoemission probing depth.
One may recognize that the Mott insulating state in Ca$_{2}$RuO$_{4}$ is also  sensitive to the excess oxygen, similarly to other perturbations~\cite{Nakatsuji2000_1,Nakatsuji2000_2,Nakamura2002,Nakamura2007,Nakamura2013}. 

Next, the energy of the incident photon, $h\nu$, was optimized before extensive ARPES measurement.
Figure~\ref{fig3}(a) shows the $h\nu$-dependence of the spectrum at the normal emission. 
The energy of the peak at $\sim -6~{\rm eV}$ changes with $h\nu$, as shown in Fig.~\ref{fig3}(b).
This is likely due to the effect of the dispersion along the perpendicular momentum, $k_{z}$, but the number of our data points were not enough to determine the value of inner potential, $V_{0}$. 
Thus, in Fig.~\ref{fig3}(b), we related $h\nu$ to $k_{z}$ by using a standard value of inner potential, $V_{0} = 10~{\rm eV}$, and a conventional formula, $k_{z} = \frac{1}{\hbar}\sqrt{2m_{\mathrm{e}}(E_{\mathrm{kin}}+V_{0})}$, where $m_{\mathrm{e}}$ is
the electron mass, and $E_{\mathrm{kin}}$ is the kinetic energy of photoelectrons.
We adopted a photon energy of $h\nu = 400~{\rm eV}$ for subsequent ARPES measurement, because the 6-eV peak has a minimum energy, likely giving a high-symmetry point in BZ, and, in relation to that, all the spectral features become sharp.\\
\hspace*{2ex}As a result, momentum-resolved electronic structure is wholly observed as shown in 
Fig.~\ref{fig4}.
The results for stoichiometric Ca$_{2}$RuO$_{4}$ are put together in the upper panels. 
Figure~\ref{fig4}(a) shows the momentum-space mapping of the oxygen band at an energy of $-7$~eV, and reveals an impressive periodic grid-like pattern.
The high-intensity lines (black) are perpendicular to $k_{x}$ and $k_{y}$ axes, and expressed as $k_{x} \simeq (0.5+n)\pi/a_\mathrm{t}$ and $k_{y} \simeq (0.5+n)\pi/a_\mathrm{t}$, where $n$ denotes an integer.
Noting that thin yellow lines indicate the unfolded tetragonal BZ, two squares around the $\Gamma$ and X points appear to be the same in size, indicating that the periodicity is in accordance with the folded orthorhombic BZ.
Figure~\ref{fig4}(b) shows that another representative momentum distribution of the oxygen band is observed at $E = -5$~eV, proving that the energy-momentum dispersion in Ca$_{2}$RuO$_{4}$ can be observed with the present experimental condition.
At $E = -5$~eV, the high-intensity lines perpendicular to the $k_{x}$ and $k_{y}$ axes cross at the $\Gamma$ point.
This image is directly comparable with the result of a previous ARPES study, and provides consistency concerning the stoichiometric samples~\cite{Sutter2017}.
Figure~\ref{fig4}(c) shows the momentum-space mapping at $E = 0$~eV.
It is quite reasonable that no significant signal is identified at $E_\mathrm{F}$, because Fig.~\ref{fig2}(b) indicated the absence of electronic states at $E_\mathrm{F}$ and the stoichiometric Ca$_{2}$RuO$_{4}$ is in the insulating phase at room temperature.\\
\hspace*{2ex}Also, the sample containing excess oxygen has been investigated under the same experimental condition.
The results are displayed in the lower panels of Fig.~\ref{fig4}, so that the counterpart is vertically aligned.
As can be seen from Figs.~\ref{fig4}(a), (b), (d), and (e), the momentum-space mappings of the oxygen band show no significant difference between the stoichiometric and excess-oxygen samples.
Concerning the momentum distribution, the oxygen band is hardly affected by a small amount of excess oxygen.
Those at $E = 0$~eV, on the other hand, present a stark contrast between two samples.
The comparison between Figs.~\ref{fig4}(c) and (f) reveals that, with excess oxygen, a new periodic square pattern emerges as a manifestation of square-shaped Fermi-surface sheets enclosing the $\Gamma$ and X points.
Taking a close look at the Fermi-surface mapping in Fig.~\ref{fig4}(f), the $\Gamma$ and X points appear to be inequivalent unlike the image plot at $E = -7$~eV in Fig.~\ref{fig4}(d).
That is why we use the tetragonal notation in the present manuscript.
It is notable that the large sheets of Fermi surface are induced by small amount of excess oxygen.
This provides evidence for the breakdown of rigid-band model for the metal-to-insulator transition of Ca$_{2}$RuO$_{4}$.\\
%
\begin{figure}[!t]
\begin{center}
\includegraphics[width=\columnwidth]{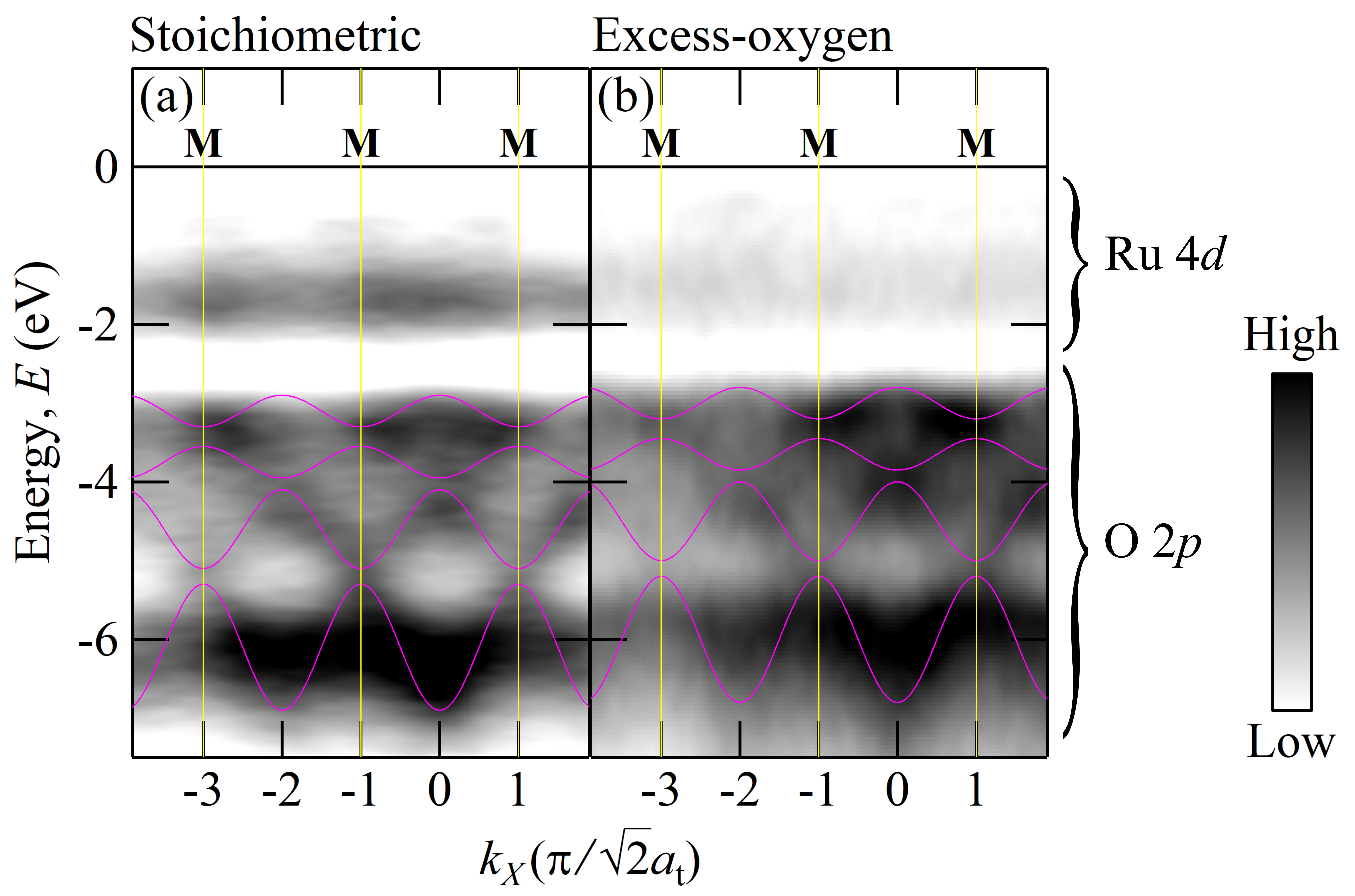}
\caption{
Dispersion of oxygen band observed along cut \#1, i.e. the diagonal M-M line cutting along the orthorhombic zone boundary.
(a) Stoichiometric sample. (b) Excess-oxygen sample. Momentum tick values are given in unit of $\pi/(\sqrt{2}a_\mathrm{t}) = 0.577$~\AA$^{-1}$, and yellow vertical lines denote the M points.
Gray scale varies from white to black following the low-to-high intensity.
Magenta sinusoids are also drawn to trace the experimental dispersion.
}
\label{fig5}
\end{center}
\end{figure}
\hspace*{2ex}The dispersion of oxygen bands is presented from another aspect in Fig.~\ref{fig5}.
The energy-momentum distribution has been extracted along the cut \#1, i.e. the diagonal M-M line cutting along the orthorhombic zone boundary as shown in Fig.~\ref{fig1}(b).
As seen from Figs.~\ref{fig5}(a) and (b), the energy of spectral peak (black area) periodically oscillates with momentum, and several dispersive bands are recognized in the energy region of the oxygen band, as traced by magenta sinusoids.
With regard to the dispersion around $-6$ eV, for example, the energy oscillates between $-5$ eV and $-7$ eV that are at the M points and their midpoints, respectively.
This feature is consistent with the momentum-space mappings in Fig.~\ref{fig4}.
In addition, the dispersion of the oxygen band shows no significant difference between stoichiometric and excess-oxygen samples.
Furthermore, the oxygen-band dispersion seen in Fig.~\ref{fig5}(a) and (b) agrees well with that predicted by density-functional-theory calculation and that previously reported from ARPES for stoichiometric Ca$_{2}$RuO$_{4}$~\cite{Sutter2017}.
Therefore, we consider that a small amount of excess oxygen hardly affects the dispersion of the oxygen bands except for the rigid-band energy shift of $\sim 0.2$ eV.\\
\hspace*{2ex}Hereafter, we focus on the Fermi surface that has emerged with introducing excess oxygen.
As seen from Fig.~\ref{fig4}(f), our observation is explained as the superposition of vertical and horizontal Fermi-surface sheets with weak hybridization.
The straight section of the Fermi surface likely arises from the orbital characters of $d_{xz}$ and $d_{yz}$, because the intersite hopping in the direction perpendicular to the lobes of the $d$ orbital is strongly suppressed.
By contrast, the Fermi surface of $d_{xy}$ band is expected to be more isotropic in the $k_{x}$-$k_{y}$ plane.
Therefore, our result is compared with a two-band model, where the $d_{xz}$ and $d_{yz}$ bands are considered as the basis functions in the present tight-binding calculation, as shown in Figs.~\ref{fig6}(a) and (b).\\
%
\hspace*{2ex}The energies in the two-band model are given as the eigenvalues of a standard $2\times 2$ Hamiltonian matrix,
\begin{displaymath}
\mathcal{H} =
\begin{pmatrix}
\varepsilon_{xz} & V\\
V & \varepsilon_{yz}\\
\end{pmatrix},
\end{displaymath}
where the diagonal terms, $\varepsilon_{xz} = -\mu-2t_{1}\cos(k_{x}a)-2t_{2}\cos(k_{y}a)$ and $\varepsilon_{yz} = -\mu-2t_{1}\cos(k_{y}a)-2t_{2}\cos(k_{x}a)$, represent the tight-binding energies of the $d_{xz}$ and $d_{yz}$ bands before hybridization, and the off-diagonal term $V = 4t_{3}\sin(k_{x}a)\sin(k_{y}a)$, provides the hybridization between two bands.
As shown in Fig.~\ref{fig1}(a), $t_{1}$ and $t_{2}$ denote the first-nearest-neighbor transfer integrals to the same $d$ orbital in the direction parallel and perpendicular to the orbital plane, respectively, $t_{3}$ denotes the second-nearest-neighbor transfer integral to the other type of $d$ orbital, and $\mu$ is the chemical potential.

The Fermi surface of the tight-binding calculation is shown in Fig.~\ref{fig6}(b).
The orange and green curves denote the lower $\alpha$ and upper $\beta$ bands, respectively, generated from the hybridization between the $d_{xz}$ and $d_{yz}$ bands as above.
The best fit with our experiment is obtained with the parameters given by $t_{2}/t_{1} = 0.08$, $t_{3}/t_{1} = 0.08$, and $\mu/t_{1} = 0.29$.
The electron occupancies of the $\alpha$ and $\beta$ bands have been determined to be $n_{\alpha} = 1.6$ and $n_{\beta} = 0.6$, respectively, from the area enclosed by the fitted Fermi surface.
The consistency between Figs.~\ref{fig6}(a) and (b) supports our simple view that only the $\alpha$ and $\beta$ sheets of the Fermi surface emerge with excess oxygen.

\begin{figure*}[!t]
\includegraphics[width=1.93\columnwidth]{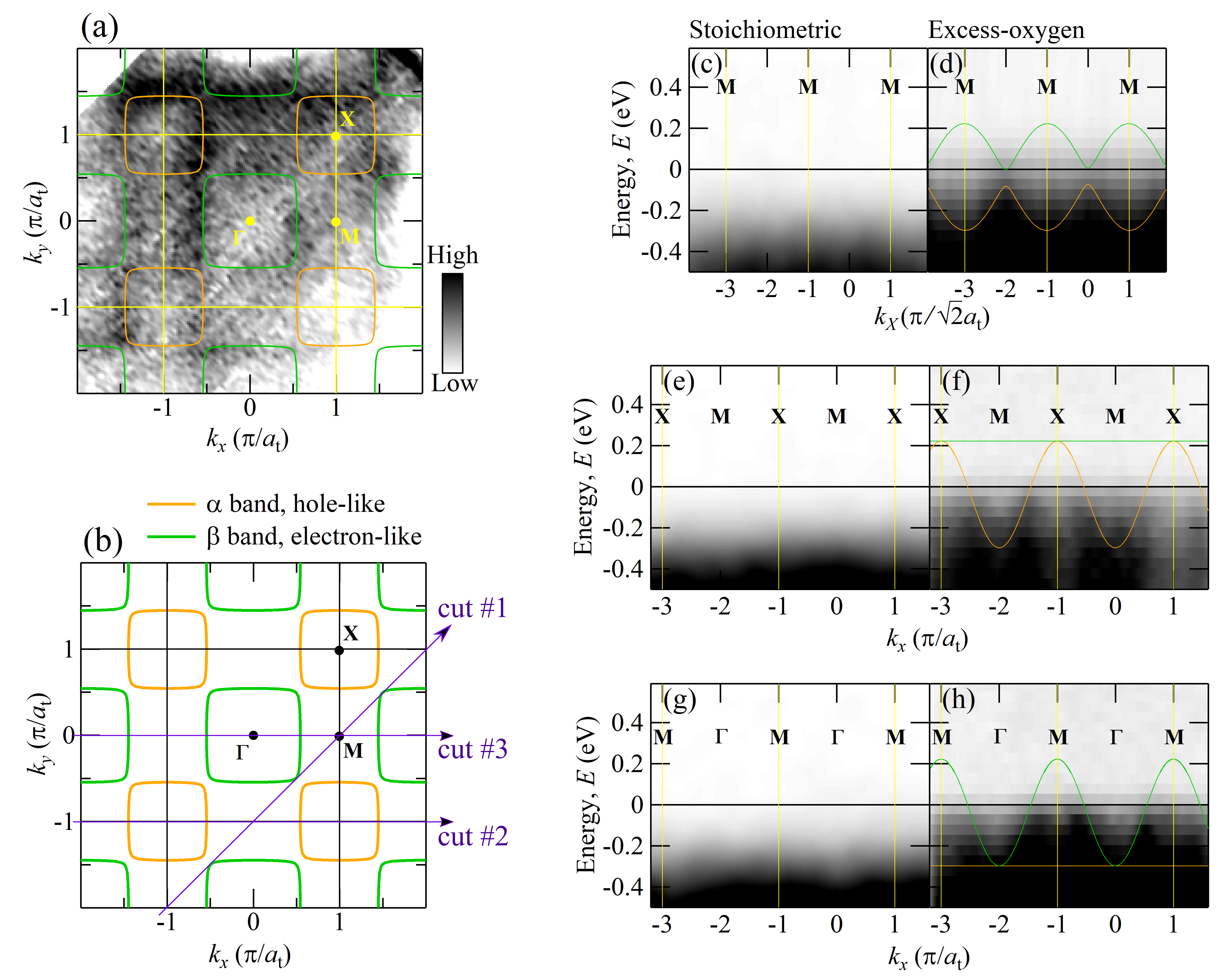}
\caption{
Low-energy electronic states of Ca$_{2}$RuO$_{4+\delta}$.
(a) Fermi-surface mapping for an excess-oxygen sample.
The image of Fig.~\ref{fig4}(f) is shown again together with the tight-binding fit.
(b) Fermi surface reproduced by tight-binding calculation with the parameters shown in the text.
Orange and green curves represent the hole-like sheet of $\alpha$ band and the electron-like sheet of $\beta$ band, respectively.
Purple arrows denote the paths of cut \#1, cut \#2, and cut \#3.
(c,d) Side-by-side comparison of ARPES intensity distributions near the Fermi level along cut \#1 for stoichiometric and excess-oxygen samples, respectively.
Gray scale varies from white to black following the low-to-high intensity.
Orange and green curves denote the dispersions from tight-binding calculation.
(e,f) Same as above but along cut \#2.
(g,h) Same as above but along cut \#3.
}
\label{fig6}
\end{figure*}

In light of the character of Fermi surface, we examine the ARPES result near $E_\mathrm{F}$.
The enlarged view of the low-energy region of Fig.~\ref{fig5} is presented in Figs.~\ref{fig6}(c) and (d).
For a stoichiometric sample, no spectral intensity is identified near $E_\mathrm{F}$, as shown in Figs.~\ref{fig6}(c), (e), and (g).
With excess oxygen, by contrast, certain spectral intensity emerges as shown in Figs.~\ref{fig6}(d), (f), and (h).
More specifically, Fig.~\ref{fig6}(d) shows that, in going along the cut \#1, the leading edge of the ARPES spectrum approaches $E_\mathrm{F}$ at the M-M midpoints and then turns to come down.
We also extracted the data along the cuts \#2 and \#3 in parallel to the $k_{x}$ axis.
Figure~\ref{fig6}(f) shows the Fermi-surface crossing of the $\alpha$ band, and reveals that the band dispersion is observed only at the M-point side of the crossing point along the cut \#2.
Figure~\ref{fig6}(h) shows the Fermi-surface crossing of the $\beta$ band, and reveals that the band dispersion is observed only at the $\Gamma$-point side of the crossing point along the cut \#3.
These results indicate that the Fermi-surface pockets around the $\Gamma$ and X points are electron-like and hole-like, respectively.
Not only the Fermi-surface mapping, but also the ARPES intensity distribution near $E_\mathrm{F}$ is described by the simple model with the $d_{xz}$ and $d_{yz}$ bands.

\section{Discussion}

Note that no folding of the Fermi surface is observed for the excess-oxygen sample, even though the bulk crystal structure is orthorhombic.
Only a single set of the electron-like and hole-like sheets of Fermi surface is observed, and no replica sheets of them are identified for the present study.
Apparently, it can be seen from Fig.~\ref{fig6} that the $\Gamma$ and X points are inequivalent to each other as if the crystal structure is tetragonal.
Indeed, it has been reported that the orthorhombic distortion is weakened upon the insulator-to-metal transition caused by excess oxygen~\cite{Braden1998}.
Because of substantial surface relaxation~\cite{Moore2007}, the surface structure is probably more sensitive to the excess oxygen than the bulk structure.
Our result suggests that the orthorhombic distortion is almost suppressed at the surface of the excess-oxygen sample.

From the viewpoint of momentum-space periodicity, the Fermi surface observed for the excess-oxygen Ca$_{2}$RuO$_{4+\delta}$ is unfolded as well as that for Sr$_{2}$RuO$_{4}$~\cite{Tamai2019}, whereas the folded Fermi surface has been observed for the metallic states induced by 10\% substitution of Sr for Ca and by uniaxial strain~\cite{Shimoyamada2009,Neupane2009,Kim2020,Ricco2018}.
Usually, small orthorhombic distortion remains after the insulator-to-metal transition of Ca$_{2}$RuO$_{4}$~\cite{Steffens2005}, and the subsequent band folding and Fermi-surface reconstruction makes intricate effects on the electronic structure~\cite{Shimoyamada2009,Neupane2009,Ricco2018,Kim2020}.
The excess oxygen, however, may provide a unique opportunity to probe the simple Fermi surface free from the band folding.

Now, we consider which band is directly involved in the insulator-to-metal transition in ruthenates.
So far, three scenarios have been proposed from the ARPES of Ca$_{1.8}$Sr$_{0.2}$RuO$_{4}$.
First, all of the $\alpha$, $\beta$, and $\gamma$ sheets of the Fermi surface were observed in Ref.~\citenum{Shimoyamada2009}.
Second, only the $\alpha$ and $\beta$ sheets were observed and the absence of $\gamma$ sheet was reported in Ref.~\citenum{Neupane2009}.
Third, the latest systematic study on Ca$_{2-x}$Sr$_{x}$RuO$_{4}$ has argued that both the $\beta$ and $\gamma$ sheets are suppressed for $x=0.2$~\cite{Kim2020}.
As another method of metallization, uniaxial strain on Ca$_{1.96}$Pr$_{0.04}$RuO$_{4}$ has been argued to induce both of the $d_{xz/yz}$ and $d_{xy}$ bands crossing at $E_\mathrm{F}$~\cite{Ricco2018}.
Our result on Ca$_{2}$RuO$_{4+\delta}$ is seemingly similar to the second scenario, because the $\alpha$ and $\beta$ sheets are evenly observed and no signal of the $\gamma$ band has been identified at $E_\mathrm{F}$, and thus indicates that the insulator-to-metal transition occurs primarily at the $\alpha$ and $\beta$ bands.

Nevertheless, the band filling provides another perspective on the evolution of Fermi surface.
It has been emphasized that the disappearance of the $\gamma$ sheet of the Fermi surface in Ca$_{1.8}$Sr$_{0.2}$RuO$_{4}$ is caused by the combination of a half-integer occupancy of the $\gamma$ band, $n_{\gamma}=1.5$, and the Fermi-surface folding into the orthorhombic BZ~\cite{Neupane2009}.
Alternatively, the proximity of the $\beta$ and $\gamma$ bands in the momentum space has been noted as a key factor for the suppression of certain sheets of the Fermi surface~\cite{Kim2020}.
One can notice from Fig.~\ref{fig6}(a) that the electron occupancies of the $\alpha$ and $\beta$ sheets for excess oxygen are somewhat lower than those reported for 10\% substitution of Sr.
Specifically, the sum of the $\alpha$- and $\beta$-band occupancies for the present study is $n_{\alpha} + n_{\beta} = 2.2$, which is appreciably lower than the value of 2.4 -- 2.51 reported for Ca$_{1.8}$Sr$_{0.2}$RuO$_{4}$~\cite{Neupane2009,Kim2020}.
The smaller filling of $\alpha$ and $\beta$ bands for Ca$_{2}$RuO$_{4+\delta}$ is reasonable, because the Sr substitution for Ca causes the upward shift of the energy of $d_{xy}$ level and thus results in the transfer of the electrons from the $\gamma$ band to the $\alpha$ and $\beta$ bands~\cite{Anisimov2002,Ko2007,Neupane2009}.
In addition to the filling, the other requirement of the orthorhombic folded BZ is also unsatisfied for the present result with excess oxygen.
It is worth noting that the primary factor in the Ca$_{2-x}$Sr$_{x}$RuO$_{4}$ system has been considered to be the bandwidth which becomes narrower at higher Ca content~\cite{Imada1998}.
Therefore, the absence of $\gamma$ sheet from Ca$_{2}$RuO$_{4+\delta}$ should be ascribed to narrower bandwidth and deeper energy level of $d_{xy}$ than for Ca$_{1.8}$Sr$_{0.2}$RuO$_{4}$.
The effects of these factors probably overcome the requirements of the band filling and band folding.
The band filling deduced from Fig.~\ref{fig6}(a) indicates that another course of orbital-selective Mott transition occurs by introducing excess oxygen in Ca$_{2}$RuO$_{4}$. \\
\hspace*{2ex}Our ARPES result involves the possible deviation between the surface and bulk properties of Ca$_{2}$RuO$_{4+\delta}$.
The present observation of the Fermi surface was rather unexpected, because the heat-capacity measurement has shown that the bulk of our sample remains in the insulating phase at room temperature even with excess oxygen.
This fact leads to an inference that $T_\mathrm{MI}$ is considerably lower at the surface than in the bulk.
Indeed, the scanning-tunneling and electron-energy-loss spectroscopic studies have revealed that $T_\mathrm{MI}$ at the surface of Ca$_{1.9}$Sr$_{0.1}$RuO$_{4}$ is lower by 24 K than that in the bulk~\cite{Moore2007}.
This downward shift of $T_\mathrm{MI}$ has been explained in relation to their discovery that the insulator-to-metal transition at the surface is accompanied by no structural transition, which stabilizes the Mott insulating phase of the bulk.
Specifically, it has been shown by low-energy electron diffraction that no abrupt lattice distortion occurs at $T_\mathrm{MI}$ for the surface of Ca$_{1.9}$Sr$_{0.1}$RuO$_{4}$, and that the elongation of RuO$_6$ octahedron along $c$-axis persists down to temperatures in the insulating phase below $T_\mathrm{MI}$ as a consequence of the broken translational symmetry at the surface~\cite{Moore2007,Moore2008}.
In addition to the surface effect, we have to consider the effect of excess oxygen, which is also known to contribute to the elongation of the RuO$_6$ octahedron along $c$-axis \cite{Nakatsuji1997,Braden1998}.
Phenomenologically and theoretically, it seems established that the $c$-axis elongation has an advantage in metallization of the ruthenates~\cite{Friedt2001,Anisimov2002,Steffens2005}.
Therefore, the effects of the surface relaxation and the excess oxygen may cooperate to decrease $T_\mathrm{MI}$.\\
\hspace*{2ex}Admittedly, it cannot be excluded that the surface inhomogeneity may enhance the discrepancy between the ARPES result and the bulk properties of Ca$_{2}$RuO$_{4+\delta}$, because the coexistence of metallic and insulating domains has been recognized for the insulator-to-metal transitions induced by current and uniaxial pressure~\cite{Zhang2019,Taniguchi2013}.
However, the insulating domains in principle make no contribution to the Fermi surface, and the distinct ARPES images demonstrate that the sufficiently large and well-crystallized region is present within the metallic domain of the excess-oxygen sample.
Thus, the electronic properties of the metallic domains of Ca$_{2}$RuO$_{4+\delta}$ are directly probed by means of the Fermi-surface mapping.

\section{Conclusions}
In conclusion, we determined the shape of Fermi surface induced by introducing excess oxygen into Ca$_{2}$RuO$_{4}$ by means of soft X-ray ARPES.
We demonstrated that eV-scale spectral-weight transfer leads to the closing of the Mott insulating gap.
We observed two square-shaped sheets, $\alpha$ and $\beta$, but no circular-shaped sheet, $\gamma$, of the Fermi surface, indicating that the Mott transition occurs selectively in the $d_{xz}$ and $d_{yz}$ bands for the oxygen-controlled Ca$_{2}$RuO$_{4+\delta}$.
However, the observed sizes of the $\alpha$ and $\beta$ sheets suggest that the absence of $\gamma$ sheet should be ascribed to narrow bandwidth and deep energy level of $d_{xy}$ for Ca$_{2}$RuO$_{4+\delta}$ unlike Ca$_{1.8}$Sr$_{0.2}$RuO$_{4}$.
Hence, our results indicate that the evolution to metallic electronic states is not unique but dependent on the method of metallization.
It follows from the finding of the unexpectedly simple Fermi surface that the systematic ARPES study of oxygen-controlled Ca$_{2}$RuO$_{4+\delta}$ is a convincing approach to uncover the nature of the Mott transition.

\section*{Acknowledgements}
We thank A. Kimura for fruitful discussions.
This work was supported by Grant-in-Aid for Scientific Research (C) (Grants Nos. 20K03842 and 19K03749) and (S) (Grants No. 17H06136).
The ARPES experiments were performed at BL25SU of SPring-8 under the approval of JASRI (Proposal No. 2018B1371 and 2019A1209). 
T. Yoshikawa were financially supported by Grants-in-Aid for JSPS Fellows No. 18J22309.


\bibliography{references}

\end{document}